# Violations of the fluctuation-dissipation theorem reveal distinct non-equilibrium dynamics of brain states

Gustavo Deco[1,2], Christopher Lynn[3,4], Yonatan Sanz Perl[1,5], and Morten L. Kringelbach[6,7,8]

1. Center for Brain and Cognition, Computational Neuroscience Group, Department of Information and Communication Technologies, Universitat Pompeu Fabra, Roc Boronat 138, Barcelona, 08018, Spain
2. Institució Catalana de la Recerca i Estudis Avançats (ICREA), Passeig Lluís Companys 23, Barcelona, 08010, Spain
3. Initiative for the Theoretical Sciences, Graduate Center, City University of New York, New York, NY 10016
4. Joseph Henry Laboratories of Physics, Princeton University, Princeton, NJ 08544
5. Department of Physics, University of Buenos Aires, Buenos Aires, Argentina
6. Centre for Eudaimonia and Human Flourishing, Linacre College, University of Oxford, Oxford, UK.
7. Department of Psychiatry, University of Oxford, Oxford, UK
8. Center for Music in the Brain, Department of Clinical Medicine, Aarhus University, Aarhus, DK.

**Abstract**

The brain is a non-equilibrium system whose dynamics change in different brain states, such as wakefulness and deep sleep. Thermodynamics provides the tools for revealing these non-equilibrium dynamics. We used violations of the fluctuation-dissipation theorem to describe the hierarchy of non-equilibrium dynamics associated with different brain states. Together with a whole-brain model fitted to empirical human neuroimaging data, and deriving the appropriate analytical expressions, we were able to capture the deviation from equilibrium in different brain states that arises from asymmetric interactions and hierarchical organisation.

Einstein and Schrödinger did not agree on many things, but they both recognised the importance of thermodynamics. At the heart of Einstein's theory of Brownian motion is the thermodynamic concept of balancing friction and thermal noise; i.e., balancing forces of dissipation and spontaneous fluctuations [1]. This balance, commonly referred to as the fluctuation-dissipation theorem (FDT), is a hallmark feature of equilibrium systems and is notably violated when systems diverge from equilibrium [2]. Later, from his exile in Ireland in the 1940s, Schrödinger came to recognise thermodynamics, and specifically the arrow of time (or irreversibility), as crucial elements for sustaining life [3]. This insight has since led to an active field of research applying non-equilibrium thermodynamics to molecular and cellular functions in systems biology, including sensing, adaptation and transportation [4-7]. Recently these ideas were extended to a thermodynamic description of whole-brain dynamics [4, 8-10], which has started to identify important changes in the hierarchical organisation and orchestration in different brain states. Specifically, by quantifying the arrow of time, one can directly measure the 'breaking of the detailed balance' in non-equilibrium systems and thereby assess the asymmetry in the flow of information. In the brain, a useful definition of hierarchy is the asymmetrical relationship between feed-forward and feed-backward interactions between brain regions. As such, a flat hierarchy is symmetric (resulting in an equilibrium system with reversible dynamics), while a hierarchical system has asymmetric interactions (resulting in irreversible dynamics that break detailed balance and diverge from equilibrium).

Here, for the first time, we use the FDT (and violations thereof) to describe the non-equilibrium dynamics associated with a given brain state. Specifically, we used a generative whole-brain model to quantify violations of the FDT in empirical neuroimaging data from human participants during different cognitive tasks, rest, and deep sleep. This perturbative model-based approach goes beyond the model-free analysis of unperturbed brain states, which cannot be used to test the FDT.

To investigate the FDT, we follow Onsager, who proposed a simple derivation using his regression principle [11-13]. This principle holds that when a system begins at an initial equilibrium state and is driven by a weak external perturbation to a final equilibrium state, the evolution of the system can be treated as a spontaneous equilibrium fluctuation. Specifically, let us assume that a weak external perturbation $\varepsilon$ is coupled to an observable $B$ at time $t = 0$. Applying Onsager's regression principle, one can derive an expression for the difference between $\langle A(t) \rangle_\varepsilon$ (the expectation value of a second observable $A$ after the perturbation is applied in $B$) and $\langle A(t) \rangle_0$ (the expectation value in the unperturbed state), which is given namely by:

$$\langle A(t) \rangle_\varepsilon - \langle A(t) \rangle_0 = \beta \varepsilon [\langle A(t)B(t) \rangle_0 - \langle A(t)B(0) \rangle_0], \qquad (1)$$

where $\beta$ is the inverse temperature from equilibrium thermodynamics. The time-dependent susceptibility is then given by:

$$\chi_{A,B}(t) = \frac{\partial \langle A(t) \rangle}{\partial \varepsilon} = \lim_{\varepsilon \to 0} \frac{\langle A(t) \rangle_\varepsilon - \langle A(t) \rangle_0}{\varepsilon} = \beta[\langle A(t)B(t) \rangle_0 - \langle A(t)B(0) \rangle_0]. \quad (2)$$

The static form of the FDT is easily obtained by taking the limit $t \to \infty$. In this case,

$$\chi_{A,B} = \beta[\langle AB \rangle_0 - \langle A \rangle_0 \langle B \rangle_0], \quad (3)$$

since correlations factorise for infinitely separated times (see Supplementary Material for a detailed derivation for spin systems). Thus, in equilibrium, we arrive at a correspondence between the response of a system to perturbation (on the left-hand side) and its unperturbed correlations (on the right-hand side).

To characterise the level of *non-equilibrium,* we can examine the normalised deviation of the system from the FDT:

$$D_{A,B} = \frac{\beta \langle AB \rangle_0 - \chi_{A,B}}{\chi_{A,B}}, \quad (4)$$

which is obtained (without loss of generality) by defining the unperturbed state such that the mean values of the observables are set to zero; i.e., $\langle A \rangle_0 = \langle B \rangle_0 = 0$. In the numerator, the first term, $\beta \langle AB \rangle_0$, corresponds to unperturbed fluctuations, while the second term, $\chi_{A,B} = \langle A \rangle_\varepsilon / \varepsilon$, corresponds to the response to a small perturbation $\varepsilon$. The total deviation $D$ can be obtained by averaging $D_{A,B}$ over all observables $A$ and all perturbation sites B. Hence, the degree of violation of the FDT, quantified by $D$, measures the divergence of the system from equilibrium. In turn, we hypothesize that these violations of the FDT will result from asymmetries in the interactions within a system, which can change from one brain state (e.g., resting versus performing a cognitive task) to another. To test this hypothesis, we investigate the spatiotemporal dynamics underlying radically different brain states using empirical human neuroimaging data recorded using functional Magnetic Resonance Imaging (fMRI).

In order to estimate the total deviation from the FDT for each participant in a given brain state, we first construct a whole-brain model fitting the corresponding functional neuroimaging data. This allows us to derive analytical expressions for the correlations between all brain regions under spontaneous fluctuations and the effect of a perturbations in one brain region on the average activities of all other regions across the brain. This whole-brain model-based analytical expression can be used to derive the total deviation from the FDT. In **Equation 4**, $D$ can be estimated after exhaustively perturbing all brain regions $B$ and observing the corresponding effects on all brain regions $A$ (see Supplementary Material for a schematic representation of the main paradigm).

To investigate the system-wide response of neural activity to targeted perturbations, we require a model of whole-brain dynamics. Here we build on the rich literature over the last ten years linking anatomical structural connectivity and functional dynamics [14-17]. The anatomical structural connectivity (SC) can be determined *in vivo* using diffusion MRI (dMRI) in conjunction with probabilistic tractography, leading to what is commonly known as the structural connectome. The whole-brain model of neural activity strikes a compromise between complexity and realism by using the physical wiring between brain regions (reflected in SC) to reproduce the empirically-measured whole-brain dynamics recorded using fMRI [17]. Such whole-brain models have had widespread success in explaining the patterns of spontaneous correlations between brain regions, forming the so-called resting-state networks [18-23].

Here, we modelled the local dynamics of each brain region as a Stuart-Landau oscillator (i.e., as the normal form of a supercritical Hopf bifurcation with bifurcation parameter, *a*), the standard model for examining the shift from noisy to oscillatory dynamics [24]. The whole-brain dynamics can be expressed by coupling the local dynamics of $N$ of these oscillators via the connectivity matrix $\boldsymbol{C}$ (see Supplementary Material for detailed explanation of the whole-brain model). Whole-brain Hopf models have been able to replicate key aspects of brain dynamics observed in electrophysiology [25, 26], magnetoencephalography [27] and fMRI [28, 29].

It has been shown that the best working point for fitting whole-brain neuroimaging dynamics is at the brink of the bifurcation, i.e. with $a_j$ slightly negative but very near to zero (usually $a_j = -0.02$, with $j = 1,..,N$ ) [30]. This proximity to criticality is crucial, because it allows a linearization of the dynamics, which, in turn, permits an analytical solution for the functional connectivity matrix $\boldsymbol{C}$, given by the Pearson correlations between all pairs of brain regions. We can estimate the functional correlations of the whole-brain network using a linear noise approximation (LNA). Hence, the dynamical system of $N$ nodes can be re-written in vector form as:

$$\frac{d\boldsymbol{z}}{dt} = (\boldsymbol{a} - \boldsymbol{S} + i\boldsymbol{\omega})\odot\boldsymbol{z} - (\boldsymbol{z}\odot\boldsymbol{\bar{z}})\boldsymbol{z} + \boldsymbol{C}\boldsymbol{z} + \boldsymbol{\eta}, \qquad (5)$$

where $\mathbf{z} = [z_1, \ldots, z_N]^T$ with $z_j = x_j + iy_j$, $\mathbf{a} = [a_1, \ldots, a_N]^T$ with $a_j$ stands for the node's bifurcation parameter, $\boldsymbol{\omega} = [\omega_1, \ldots, \omega_N]^T$ where $\omega_j$ is the intrinsic node frequency, $\boldsymbol{\eta} = [\eta_1, \ldots, \eta_N]^T$ where $\eta_j$ is additive uncorrelated Gaussian noise with variance $\sigma^2$ (for all $j$), and $\mathbf{S} = [S_1, \ldots, S_N]^T$ is a vector containing the connectivity strength of each node; i.e., $S_i = \sum_j C_{ij}$. The superscript $[\ ]^T$ represents the transpose, $\odot$ is the Hadamard element-wise product, and $\bar{\mathbf{z}}$ is the complex conjugate of $\mathbf{z}$. This equation describes the linear fluctuations around the fixed point $\mathbf{z} = 0$, which is the solution of $\frac{d\mathbf{z}}{dt} = 0$. Separating the real and imaginary parts of the state variables, and discarding the higher-order terms $(\mathbf{z} \odot \bar{\mathbf{z}})\mathbf{z}$, the evolution of the linear fluctuations follows a Langevin stochastic linear equation:

$$\frac{d}{dt} \delta \mathbf{u} = \mathbf{J} \delta \mathbf{u} + \boldsymbol{\eta}, \tag{6}$$

where the 2$N$-dimensional vector $\delta \mathbf{u} = [\delta \mathbf{x}, \delta \mathbf{y}]^T = [\delta x_1, \ldots, \delta x_N, \delta y_1, \ldots, \delta y_N]^T$ contains the fluctuations of real and imaginary state variables. The $2N \times 2N$ matrix $\mathbf{J}$ is the Jacobian of the system evaluated at the fixed point, which can be written as a block matrix

$$\mathbf{J} = \begin{bmatrix} \mathbf{J}_{xx} & \mathbf{J}_{xy} \\ \mathbf{J}_{yx} & \mathbf{J}_{yy} \end{bmatrix}, \tag{7}$$

where $\mathbf{J}_{xx}$, $\mathbf{J}_{xy}$, $\mathbf{J}_{yx}$, $\mathbf{J}_{yy}$ are $N \times N$ matrices $\mathbf{J}_{xx} = \mathbf{J}_{yy} = \text{diag}(\mathbf{a} - \mathbf{S}) + \mathbf{C}$ and $\mathbf{J}_{xy} = -\mathbf{J}_{yx} = \text{diag}(\boldsymbol{\omega})$, where $\text{diag}(\mathbf{v})$ is the diagonal matrix whose diagonal is the vector $\mathbf{v}$. We note that the above linearization is only valid if $\mathbf{z} = 0$ is a stable solution of the system; i.e., if all eigenvalues of $\mathbf{J}$ have negative real part.

To examine the FDT, we must first compute the covariance matrix $\mathbf{K} = \langle \delta \mathbf{u} \delta \mathbf{u}^T \rangle$. We begin by writing **Equation 6** as $d \delta \mathbf{u} = \mathbf{J} \delta \mathbf{u} dt + d\mathbf{W}$, where $d\mathbf{W}$ is an 2$N$-dimensional Wiener process with covariance $\langle d\mathbf{W} d\mathbf{W}^T \rangle = \mathbf{Q} dt$ and $\mathbf{Q}$ is the noise covariance matrix (which is diagonal if the noise is uncorrelated). Using Itô's stochastic calculus, we get $d(\delta \mathbf{u} \delta \mathbf{u}^T) = d(\delta \mathbf{u}) \delta \mathbf{u}^T + \delta \mathbf{u} d(\delta \mathbf{u}^T) + d(\delta \mathbf{u}) d(\delta \mathbf{u}^T)$. Taking expectations, keeping terms to first order in the differential $dt$, and noting that $\langle \delta \mathbf{u} d\mathbf{W}^T \rangle = 0$, we obtain:

$$\frac{d\mathbf{K}}{dt} = \mathbf{J} \mathbf{K} + \mathbf{K} \mathbf{J}^T + \mathbf{Q}. \tag{8}$$

Hence, the stationary covariances can be obtained by analytically solving the **Equation 8** for the case $\frac{dK}{dt} = 0$.

This Lyapunov equation can be solved using the eigen-decomposition of the Jacobian matrix $J$ [31]. We then obtained the simulated functional connectivity $FC^{model}$ from the first $N$ rows and columns of the covariance $K$, which corresponds to the real part of the dynamics (precisely representing the BOLD fMRI signal).

Still, even if the analytical solution is possible, in order to fit the model to the empirical data (BOLD fMRI of each participant in each brain state), for the optimization of the coupling connectivity matrix $C$, similar to the work of Gilson and colleagues, here it proved more robust to estimate this numerically by using a pseudo-gradient descent procedure [32, 33]. Specifically, we fit $C$ (see Supplementary Material) such that the model optimally reproduces the empirically measured covariances $FC^{empirical}$ (i.e., the normalised covariance matrix of the functional neuroimaging data) and the empirical time-shifted covariances $FS^{empirical}(\tau)$, where $\tau$ is the time lag, which are normalized for each pair of regions $i$ and $j$ by $\sqrt{KS_{ii}^{empirical}(0)KS_{jj}^{empirical}(0)}$. We selected the parameter $\tau$, which led to a decrease in the averaged autocorrelation. We note that fitting the time-shifted correlations can lead to asymmetries in the connectivity $C$, which, in turn, can produce non-equilibrium dynamics and violations of the FDT. These normalised time-shifted covariance matrices are generated by taking the shifted covariance matrix $KS^{empirical}(\tau)$ and dividing each pair $(i, j)$ by $\sqrt{KS_{ii}^{empirical}(0)KS_{jj}^{empirical}(0)}$. Note that these normalised time-shifted covariances break the symmetry of the couplings and thus improve the level of fitting [34]. Using a heuristic pseudo-gradient algorithm, we proceeded to update the $C$ until the fit is fully optimised. More specifically, the updating uses the following form:

$$C_{ij} = C_{ij} + \alpha\left(FC_{ij}^{empirical} - FC_{ij}^{model}\right) + \varsigma\left(FS_{ij}^{empirical}(\tau) - FS_{ij}^{model}(\tau)\right), \qquad (9)$$

where $FS_{ij}^{model}(\tau)$ is defined similar to $FS_{ij}^{empirical}(\tau)$. In other words it is given by the first $N$ rows and columns of the simulated $\tau$ time-shifted covariances $KS^{model}(\tau)$ normalized by dividing each pair $(i, j)$ by $\sqrt{KS_{ii}^{model}(0)KS_{jj}^{model}(0)}$, being $\boldsymbol{KS}^{model}(\tau)$ the shifted simulated covariance matrix computed as following:

$$\boldsymbol{KS}^{model}(\tau) = \exp(\tau \boldsymbol{J})\, \boldsymbol{K}. \tag{10}$$

Note that $\boldsymbol{KS}^{model}(0) = \boldsymbol{K}$. The model was run repeatedly with the updated $\boldsymbol{C}$ until the fit converges towards a stable value. We initialised $\boldsymbol{C}$ using the anatomical connectivity (obtained with probabilistic tractography from dMRI) and only update known existing connections from this matrix (in either hemisphere). However, there is one exception to this rule which is that the algorithm also updates homologue connections between the same regions in either hemisphere, given that tractography is known to be less accurate when accounting for this connectivity. For the Stuart-Landau model, we used $\alpha = \varsigma = 0.00001$ and continue until the algorithm converges. For each iteration we compute the model results as the average over as many simulations as there are participants. Overall, we use the term Generative Effective Connectivity (GEC) for the optimised $\boldsymbol{C}$ [35].

After fitting an individualized coupling matrix $\boldsymbol{C}$ for each participant and each brain state, we derived an analytical form for the deviation from the FDT (corresponding to **Equation 4**). First, we derive the expectation values of the state variables $\langle \delta \boldsymbol{u} \rangle_{\varepsilon_j}$ when a perturbation $\varepsilon$ is applied to the component $j$. From **Equation 6**, we have the relationship $\frac{d}{dt}\langle \delta \boldsymbol{u} \rangle_{\varepsilon_j} = \boldsymbol{J}\langle \delta \boldsymbol{u} \rangle_{\varepsilon_j} + \boldsymbol{h}_j = 0$, where $\boldsymbol{h}_j$ is a $2N$-dimensional vector of all zeros except for the $j$ component, which is equal to $\varepsilon$. Solving for the desired expectation value, we obtain $\langle \delta \boldsymbol{u} \rangle_{\varepsilon_j} = -\boldsymbol{J}^{-1}\boldsymbol{h}_j$. Defining $\langle \delta \boldsymbol{x} \rangle_j = \langle \delta \boldsymbol{x} \rangle_{\varepsilon_j}/\varepsilon$, i.e. the real part of $\langle \delta \boldsymbol{u} \rangle_j$, we can now derive the deviation from the FDT for region $i$ when a perturbation is applied to region $j$:

$$D_{i,j} = \frac{2\langle \delta x_i \delta x \rangle_0 / \sigma^2 - \langle \delta x_i \rangle_j}{\langle \delta x_i \rangle_j}, \tag{11}$$

where the term $2/\sigma^2$ plays the role of the inverse temperature $\beta$, and the covariance $\langle \delta x_i \delta x \rangle_0$ is derived from $\boldsymbol{KS}^{model}$. For numerical reasons, we quantify the system-wide effect of perturbing the component $j$ by averaging the numerator and denominator over the regions; i.e.,

$$P_j = \frac{\frac{1}{N}\sum_i 2\langle \delta x_i \delta x_j \rangle_0 / \sigma^2 - \langle \delta x_i \rangle_j}{\frac{1}{N}\sum_i \langle \delta x_i \rangle_j}. \tag{12}$$

The vector $P$ defines a *perturbability map* over all brain regions in a given brain state. For each participant, the level of non-equilibrium $\widehat{D}$ is finally computed by averaging the deviation from the

FDT over all possible perturbations, i.e., $\widehat{D} = \frac{1}{N}\sum_j P_j$. We applied this FDT framework to two empirical neuroimaging datasets in humans, with whole-brain activity measured using BOLD fMRI. The first dataset consists of 18 human participants whose sleep stages were precisely characterised by two independent neurologists from simultaneous electroencephalography (EEG) recordings [36]. We considered two stages of consciousness: wakefulness and deep sleep (N3) (see Supplementary Material for details on the experimental setup and data processing). The second dataset consists of 970 participants from the Human Connectome Project (again, see Supplementary Material for details), who were recorded during resting state and seven different tasks spanning a broad range of cognitive and emotional processing [37].

First, we applied the FDT framework to the sleep dataset and found significant differences in the deviations from the FDT when comparing deep sleep with wakefulness (p<0.001, permutation test). Specifically, we computed $\widehat{D}$ for each participant and each level of consciousness, revealing a decrease in violations of the FDT (or level of non-equilibrium) during deep sleep compared with wakefulness. This difference can be interpreted as a flattening of the hierarchical organisation during deep sleep; that is, a brain state with more symmetrical interactions compared with wakefulness. These violations of the FDT can be clearly visualized using the corresponding perturbability maps (vector $P_j$ from **Equation 12**), which show more homogeneous and much lower levels of non-equilibrium responses for deep sleep compared to wakefulness.

Second, we observed significant differences in the violations of the FDT when comparing resting state with different cognitive tasks across 970 healthy participants (p<0.001 for all comparisons, permutation tests). Just as in the investigations of sleep states, we computed $\widehat{D}$ for each participant and each cognitive task (including rest), revealing differences in the non-equilibrium nature of the brain. For example, the SOCIAL task induced the largest violations of the FDT, reflecting the highest level of non-equilibrium [4, 9, 10]. By contrast, we observed closer agreement with the FDT for resting compared to each of the cognitive tasks. Indeed, the perturbability map for rest is more homogeneous with responses that are closer to equilibrium compared to the SOCIAL task. These results suggest that violations of the FDT (and the distance from equilibrium) increase with computational demands. This can be interpreted in terms of the breaking of the detailed balance, where the flow of information requires asymmetric interactions between brain regions.

To investigate the mechanisms underlying violations of the FDT, and the relationship to the breaking of detailed balance, we constructed two simple linear models with differing levels of asymmetry in their interactions. Specifically, to relate the asymmetry of the underlying coupling matrix to violations of the FDT, we use a Langevin equation ($\frac{d\boldsymbol{b}}{dt} = \boldsymbol{L}\boldsymbol{b} + \boldsymbol{\eta}$), where $\boldsymbol{b} = [b_1, \ldots, b_N]^T$ models the bold signal in a parcellation of N regions, $\boldsymbol{L}$ is the coupling matrix, and $\boldsymbol{\eta} = [\eta_1, \ldots, \eta_N]^T$ the additive Gaussian noise. We consider two different models, each generated by fitting the couplings $\boldsymbol{L}$ to the empirical neural activity during wakefulness to obtain realistic generative effective connectivity. We first define a symmetric model that is only fit to the equal-time empirical correlations, resulting in symmetric effective connectivity $\boldsymbol{L}$. We then define an asymmetric model that fit to both the equal-time and time-delayed correlations, resulting in asymmetric connectivity.

**Figure 2** shows the importance of asymmetric couplings for violations of the FDT. Specifically, **Figure 2A** shows how the linear *symmetric* model generates a fully symmetric connectivity matrix (left panel), which can be observed by computing $|\boldsymbol{L} - \boldsymbol{L}^T|$ (middle panel). Notably, these symmetric couplings yield fully equilibrium dynamics, and therefore do not generate any violations of the FDT (right panel).

By contrast, **Figure 2B** show the connectivity matrix of the asymmetric model (left panel), which is asymmetric (middle panel) and thus induces significant violations of the FDT (right panel). These differences between the asymmetric and symmetric models are illustrated in **Figure 2C**. The first panel shows a scatterplot of the mean regional FDT deviations (i.e., averaging over the rows of the deviation matrix $D_{i,j}$) as a function of the mean regional connectivity strength (with red points for the asymmetrical model and black points for the symmetrical model). In the asymmetric model, we observe a significant correlation (of 0.77) between the FDT deviations and the regional connectivity strength, while this correlation vanishes for the symmetric model. The second panel shows the scatterplot of the mean perturbation site FDT deviation (i.e., averaging over the columns of the deviation matrix $D_{i,j}$) as a function of the mean perturbation site connectivity strength, revealing a negative correlation (-0.87) for the asymmetric model. The third panel shows a violin plot of the significant mean FDT deviation for the symmetric (grey) and asymmetric (green) models across all regions and sites (p<0.001, permutation testing).

**Discussion**

For the first time, we applied the FDT to neural activity fitted by a whole-brain model, which allowed us to investigate how non-equilibrium dynamics are associated with sleep, wakefulness and seven cognitive tasks. We find that violations of the FDT (and thus divergences from equilibrium) are driven by asymmetries in the couplings between brain regions, thus revealing the role of hierarchical organization in non-equilibrium dynamics. The largest violations of the FDT were observed when subjects performed cognitive tasks (with the SOCIAL task inducing the largest violations), while the neural dynamics were closer to equilibrium for sleep than wakefulness. These differences directly reflect the computational demands that require asymmetric information flow between brain regions, thus breaking detailed balance and promoting non-equilibrium dynamics. Schrödinger hypothesized that this increasing asymmetrical information flow is important for sustaining life [3], and here we extend this thermodynamic principle to neural computations.

Using thermodynamics to describe brain dynamics is an emerging field [21], which has already yielded important new insights into the non-equilibrium nature of brain function [4, 8-10]. Excitingly, these insights include the demonstration of how the arrow of time, or irreversibility, of brain signals can shed new light on the definition of brain states [9, 10, 35]. Meanwhile, brain dynamics have also been shown to be turbulent [38, 39], allowing the fast information transfer needed for time-critical decisions in the brain (MEG).

Overall, the model-based FDT approach introduced here holds great promise for revealing the underlying principles of non-equilibrium dynamics in the human brain. Specifically, this approach may shed new light on the changes in hierarchical organisation and asymmetric interactions in health and, perhaps more importantly, the breakdown in neuropsychiatric disease.

## Acknowledgements

G.D. was supported by the Human Brain Project Specific Grant Agreement 3 Grant agreement no. 945539 and by the Spanish Research Project AWAKENING: using whole-brain models perturbational approaches for predicting external stimulation to force transitions between different brain states, ref. PID2019-105772GB-I00/AEI/10.13039/501100011033, financed by the Spanish Ministry of Science, Innovation and Universities (MCIU), State Research Agency (AEI). C.W.L. is supported by a James S. McDonnell Foundation Postdoctoral Fellowship Award and by the National Science Foundation through the Center for the Physics of Biological Function (PHY–1734030). Y.S.P is supported by European Union's Horizon 2020 research and innovation programme under the Marie Sklodowska-Curie grant 896354. M.L.K. is supported by the Center for Music in the Brain, funded by the Danish National Research Foundation (DNRF117), and Centre for Eudaimonia and Human Flourishing at Linacre College funded by the Pettit and Carlsberg Foundations.

# Figures

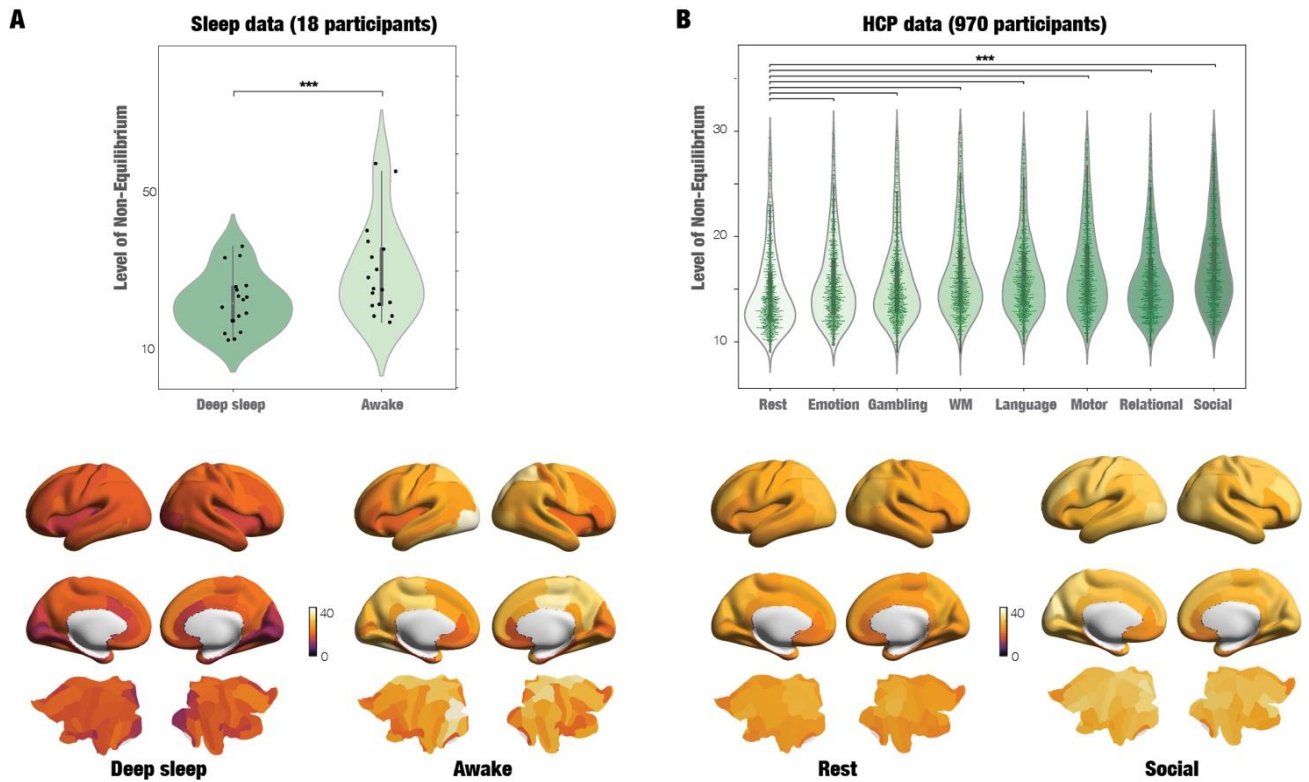

*Figure 1. Non-equilibrium fingerprints of brain states captured by deviations from FDT. A) Significant differences in deviations from FDT were found when comparing deep sleep with wakefulness in neuroimaging data from 18 healthy participants. Renderings of the resulting perturbability maps on the human brain show more homogeneous and much lower levels of non-equilibrium for deep sleep compared to wakefulness. B) Similarly, significant differences were observed when comparing resting state with seven different tasks in 970 participants from the human connectome project (p<0.001 for all comparisons). As can be seen the perturbability maps are more homogeneous and much lower levels of non-equilibrium for rest compared to task (here for the SOCIAL task).*

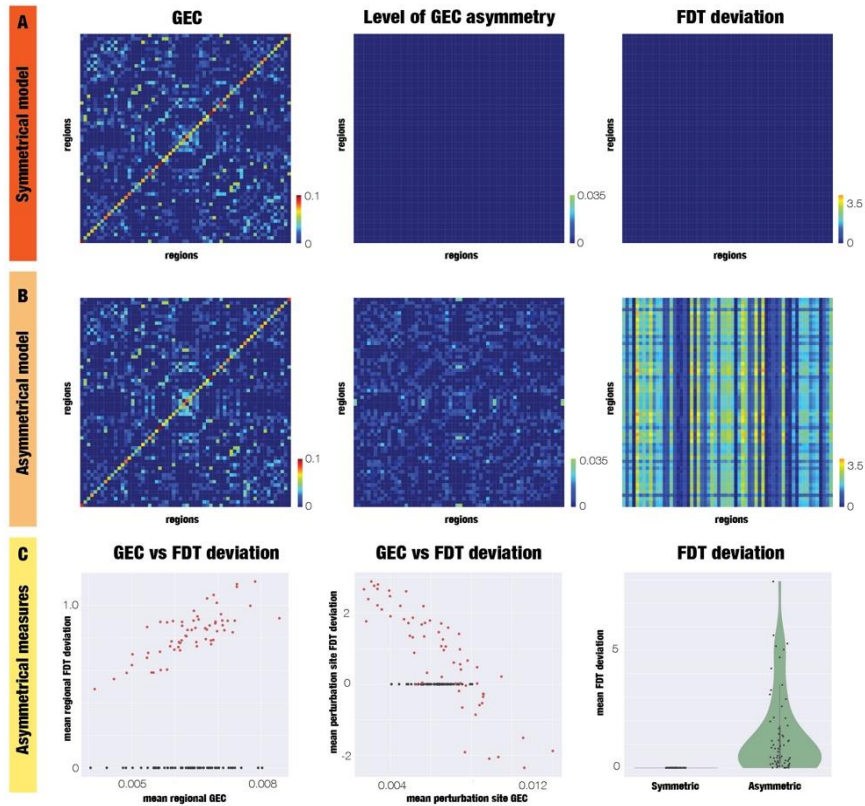

*Figure 2. Whole-brain models show the importance of asymmetric coupling compared to symmetric coupling.* **A)** *The simplest linear symmetric model generates a Generative Effective Connectivity (GEC) matrix (shown on the left) which is fully symmetrical (shown by the middle matrix) and does not generate any deviations from FDT as shown by the right matrix (level of FDT deviation).* **B)** *The asymmetric model generates a GEC matrix (left) which is asymmetric (middle) and with significant deviations from FDT (right).* **C)** *The first panel shows a scatterplot of the mean regional FDT deviation as a function of the mean regional GEC. The second panel shows a scatterplot of the mean perturbation site FDT deviation as a function of the mean perturbation site GEC. In both cases the asymmetrical model (red points) shows correlation between FDT deviation and mean GEC metrics in contrast with the symmetrical model (black points). Finally, the right panel shows a violin plot of the mean FDT deviation for the symmetrical (grey) and asymmetrical (green) models across all regions and sites.*

# Supplementary Material for: "Violations of the fluctuation-dissipation theorem reveal distinct non-equilibrium dynamics of brain states"

Gustavo Deco[1,2], Christopher Lynn[3,4], Yonatan Sanz Perl[1,5], and Morten L. Kringelbach[6,7,8]

1. Center for Brain and Cognition, Computational Neuroscience Group, Department of Information and Communication Technologies, Universitat Pompeu Fabra, Roc Boronat 138, Barcelona, 08018, Spain
2. Institució Catalana de la Recerca i Estudis Avançats (ICREA), Passeig Lluís Companys 23, Barcelona, 08010, Spain
3. Initiative for the Theoretical Sciences, Graduate Center, City University of New York, New York, NY 10016
4. Joseph Henry Laboratories of Physics, Princeton University, Princeton, NJ 08544
5. Department of Physics, University of Buenos Aires, Buenos Aires, Argentina
6. Centre for Eudaimonia and Human Flourishing, Linacre College, University of Oxford, Oxford, UK.
7. Department of Psychiatry, University of Oxford, Oxford, UK
8. Center for Music in the Brain, Department of Clinical Medicine, Aarhus University, Aarhus, DK.

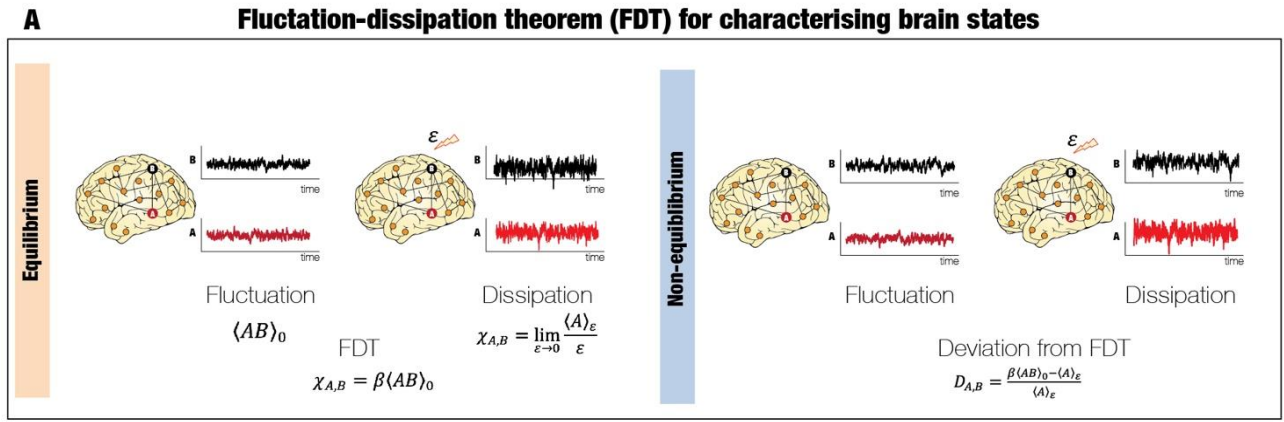

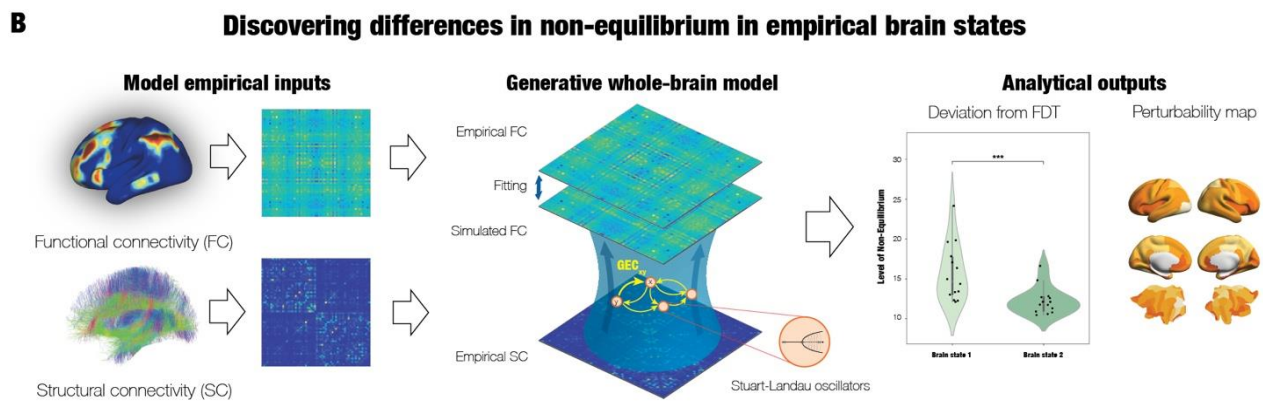

*Figure S1. Fluctuation-dissipation theorem (FDT) used on empirical neuroimaging data. A)* As can be seen from the general framework of FDT in equilibrium (left panel) and non-equilibrium (right panel), this can be used to characterise different brain states. Specifically, the level of non-equilibrium can be captured as the deviation of FDT and can subsequently be used to describe the orchestration and changes in hierarchy. *B)* Combining FDT with a whole-brain model (linking anatomical connectivity and functional brain connectivity) fitted to empirical neuroimaging data can precisely describe the overall deviation from FDT as well as the perturbability maps for different brain states.

*Stuart-Landau Whole-brain model*

Given a parcellation of $N$ regions, the whole-brain dynamics can be expressed by coupling the local dynamics of $N$ Stuart-Landau oscillators via the connectivity matrix $\mathbf{C}$, which is defined by

$$\frac{dz_j}{dt} = (a_j + i\omega_j)z_j - |z_j|^2 z_j + \sum_{k=1}^{N} C_{jk}(z_k - z_j) + \eta_j, \qquad (5)$$

where the complex variable $z_j$ denotes the state ($z_j = x_j + iy_j$) of region $j$, $\eta_j$ is additive uncorrelated Gaussian noise with variance $\sigma^2$ (for all $j$), $\omega_j$ is the intrinsic node frequency, and $a_j$ is the node's bifurcation parameter. The intrinsic frequencies $\omega_j$ (which lie in the 0.008–0.08Hz band) were estimated from the data as the averaged peak frequencies of the narrowband blood-oxygen-level-dependent (BOLD) signals of the different brain regions. For $a_j > 0$, the local dynamics settle into a stable limit cycle, producing self-sustained oscillations with frequency $\omega_j/(2\pi)$. For $a_j < 0$, the local dynamics present a stable spiral point, producing damped or noisy oscillations in the absence or presence of noise, respectively. The fMRI signals were modelled by the real part of the state variables; i.e., $x_j = \text{Real}(z_j)$.

*Parcellation*

Both datasets used timeseries from the Mindboggle-modified Desikan-Killiany parcellation[1] with a total of 62 cortical regions (31 regions per hemisphere)[2].

*Human Connectome project: Acquisition and pre-processing*

*Ethics*

The Washington University–University of Minnesota (WU-Minn HCP) Consortium obtained full informed consent from all participants, and research procedures and ethical guidelines were followed in accordance with Washington University institutional review board approval (Mapping the Human Connectome: Structure, Function, and Heritability; IRB # 201204036).

*Participants*

The data set used for this investigation was selected from the March 2017 public data release from the Human Connectome Project (HCP) where we chose a sample of 1003 participants, all of whom have resting state data. For the seven tasks, HCP provides the following numbers of participants: WM =999; SOCIAL=996; MOTOR=996; LANGUAGE=997; GAMBLING=1000; EMOTION=992;

RELATIONAL=989. No statistical methods were used to pre-determine sample sizes but our sample sizes are similar to those reported in previous publications using the full HCP dataset.

*The HCP task battery of seven tasks*

The HCP task battery consists of seven tasks: working memory, motor, gambling, language, social, emotional, relational, which are described in details on the HCP website. HCP states that the tasks were designed to cover a broad range of human cognitive abilities in seven major domains that sample the diversity of neural systems 1) visual, motion, somatosensory, and motor systems, 2) working memory, decision-making and cognitive control systems; 3) category-specific representations; 4) language processing; 5) relational processing; 6) social cognition; and 7) emotion processing. In addition to resting state scans, all 1003 HCP participants performed all tasks in two separate sessions (first session: working memory, gambling and motor; second session: language, social cognition, relational processing and emotion processing).

*3T structural data*

The HCP structural data were acquired using a customized 3 Tesla Siemens Connectom Skyra scanner with a standard Siemens 32-channel RF-receive head coil. For each participant, at least one 3D T1w MPRAGE image and one 3D T2w SPACE image were collected at 0.7 mm isotropic resolution.

*3T diffusion MRI*

In order to reconstruct a high-quality structural connectivity (SC) matrix for constructing the whole-brain model (using the DK62 parcellation), we obtained multi-shell diffusion-weighted imaging data from 32 participants from the HCP database (scanned for approximately 89 minutes). The acquisition parameters are described in detail on the HCP website[3]. We estimated the connectivity using the method described by Horn and colleagues[4]. Briefly, the data was processed using a generalized q-sampling imaging algorithm implemented in DSI studio (http://dsi-studio.labsolver.org). Segmentation of the T2-weighted anatomical images produced a white-matter mask and co-registering the images to the b0 image of the diffusion data using SPM12. In each HCP participant, 200,000 fibres were sampled within the white-matter mask. Fibres were transformed into MNI space using Lead-DBS[5]. The methods used the algorithms for false-positive fibres shown to be optimal in recent open challenges [6,7]. The risk of false positive tractography was reduced in several ways. Most importantly, this used the tracking method achieving the highest (92%) valid connection score among 96 methods submitted from 20 different research groups in a recent open competition[6].

*Neuroimaging acquisition for fMRI HCP*

The 1003 HCP participants were scanned on a 3-T connectome-Skyra scanner (Siemens). We used one resting state fMRI acquisition of approximately 15 minutes acquired on the same day, with eyes open with relaxed fixation on a projected bright cross-hair on a dark background as well as data from the seven tasks. The HCP website (http://www.humanconnectome.org/) provides the full details of participants, the acquisition protocol and pre-processing of the data for both resting state and the seven tasks. Below we have briefly summarised these.

The pre-processing of the HCP resting state and task datasets is described in details on the HCP website. Briefly, the data is pre-processed using the HCP pipeline which is using standardized methods using FSL (FMRIB Software Library), FreeSurfer, and the Connectome Workbench software[8,9]. This standard pre-processing included correction for spatial and gradient distortions and head motion, intensity normalization and bias field removal, registration to the T1 weighted structural image, transformation to the 2mm Montreal Neurological Institute (MNI) space, and using the FIX artefact removal procedure [9,10]. The head motion parameters were regressed out and structured artefacts were removed by ICA+FIX processing (Independent Component Analysis followed by FMRIB's ICA-based X-noiseifier[11,12]). Pre-processed timeseries of all grayordinates are in HCP CIFTI grayordinates standard space and available in the surface-based CIFTI file for each participants for resting state and each of the seven tasks.

We used a custom-made Matlab script using the ft_read_cifti function (Fieldtrip toolbox [13]) to extract the average timeseries of all the grayordinates in each region of the Mindboggle-modified Desikan-Killiany parcellation[1] with a total of 62 cortical regions (31 regions per hemisphere) [2], which are defined in the HCP CIFTI grayordinates standard space. The BOLD timeseries were filtered using a second-order Butterworth filter in the range of 0.008-0.08Hz.

*Human sleep data: Acquisition and pre-processing*

*Ethics*

Written informed consent was obtained, and the study was approved by the ethics committee of the Faculty of Medicine at the Goethe University of Frankfurt, Germany.

*Participants*

We used fMRI- and PSG data from 18 participants taken from a larger database that reached all four stages of PSG[14]. Exclusion criteria focussed on the quality of the concomitant acquisition of EEG, EMG, fMRI, and physiological recordings.

*Acquisition and pre-processing of fMRI and polysomnography data*

Neuroimaging fMRI was acquired on a 3 T system (Siemens Trio, Erlangen, Germany) with the following settings: 1505 volumes of T2*-weighted echo planar images with a repetition time (TR) of 2.08 seconds, and an echo time of 30 ms; matrix 64 x 64, voxel size 3 x 3 x 2 mm$^3$, distance factor 50%, FOV 192 mm$^2$.

The EPI data were realigned, normalised to MNI space, and spatially smoothed using a Gaussian kernel of 8 mm$^3$ FWHM in SPM8 (http://www.fil.ion.ucl.ac.uk/spm/). Spatial downsampling was then performed to a 4 x 4 x 4 mm resolution. From the simultaneously recorded ECG and respiration, cardiac- and respiratory-induced noise components were estimated using the RETROICOR method [15], and together with motion parameters these were regressed out of the signals. The data were temporally band-pass filtered in the range 0.008-0.08 Hz using a sixth-order Butterworth filter. We extracted the timeseries in the DK62 parcellation [16].

Simultaneous PSG was performed through the recording of EEG, EMG, ECG, EOG, pulse oximetry, and respiration. EEG was recorded using a cap (modified BrainCapMR, Easycap, Herrsching, Germany) with 30 channels, of which the FCz electrode was used as reference. The sampling rate of the EEG was 5 kHz, and a low-pass filter was applied at 250 Hz. MRI and pulse artefact correction were applied based on the average artefact subtraction method [17] in Vision Analyzer2 (Brain Products, Germany). EMG was collected with chin and tibial derivations, and as the ECG and EOG recorded bipolarly at a sampling rate of 5 kHz with a low-pass filter at 1 kHz. Pulse oximetry was collected using the Trio scanner, and respiration with MR-compatible devices (BrainAmp MR+, BrainAmp ExG; Brain Products, Gilching, Germany).

Participants were instructed to lie still in the scanner with their eyes closed and relax. Sleep classification was performed by a sleep expert based on the EEG recordings in accordance with the AASM criteria (2007). Results using the same data and the same pre-processing has previously been reported [14].

*Statistical comparisons*

Differences in probabilities of occurrence before and after injection were statistically assessed using a permutation-based paired t-test. This non-parametric test uses permutations of group labels to estimate the null distribution, which is computed independently for each experimental condition. For each of 1,000 permutations, a t-test is applied to compare populations and a p-value is returned.

*Derivation of the fluctuation-dissipation theorem in spin systems*

The simplest thermodynamic model of a system with multiple interacting components is the Ising model. In the Ising model, each component (or spin) is represented by a binary variable $x_i$. The probability of finding the entire system in state $x = \{x_i\}$ is given by the Boltzmann distribution

$$P(x) = \frac{1}{Z}\exp\left[\beta\left(\sum_{i,j} J_{ij} x_i x_j + \sum_i h_i x_i\right)\right], \tag{S1}$$

Where $\beta$ is the inverse temperature, $J_{ij} = J_{ji}$ represents the strength of the interaction between components $i$ and $j$, $h_i$ is the external influence on component $i$, and

$$Z = \sum_x \exp\left[\beta\left(\sum_{i,j} J_{ij} x_i x_j + \sum_i h_i x_i\right)\right] \tag{S2}$$

is the normalization constant (often referred to as the partition function). Because the interactions $J_{ij}$ are symmetric, the system is in equilibrium and the fluctuation-dissipation theorem should hold.

To derive the fluctuation-dissipation theorem, we would like to know how the average state of component $i$

$$\langle x_i \rangle = \sum_x x_i P(x) \tag{S3}$$

changes due to a small perturbation $h_j$ coupled to component $j$. In particular, we have

$$\chi_{ij} = \frac{\partial \langle x_i \rangle}{\partial h_j} = \sum_x x_i \frac{\partial}{\partial h_j} P(x) \tag{S4}$$

$$= \frac{1}{Z}\sum_x x_i \frac{\partial}{\partial h_j}\exp\left[\beta\left(\sum_{k,l} J_{kl} x_k x_l + \sum_k h_k x_k\right)\right] - \frac{1}{Z^2}\frac{\partial Z}{\partial h_j}\sum_x x_i \exp\left[\beta\left(\sum_{k,l} J_{kl} x_k x_l + \sum_k h_k x_k\right)\right].$$

For the first term, we have

$$\frac{1}{Z}\sum_x x_i \frac{\partial}{\partial h_j}\exp\left[\beta\left(\sum_{k,l} J_{kl} x_k x_l + \sum_k h_k x_k\right)\right] = \frac{\beta}{Z}\sum_x x_i x_j \exp\left[\beta\left(\sum_{k,l} J_{kl} x_k x_l + \sum_k h_k x_k\right)\right] \tag{S5}$$

$$= \beta \sum_x x_i x_j P(x) = \beta \langle x_i x_j \rangle.$$

For the second term, we first note that

$$\frac{1}{Z}\frac{\partial Z}{\partial h_j} = \frac{1}{Z}\sum_x \frac{\partial}{\partial h_j}\exp\left[\beta\left(\sum_{k,l}J_{kl}x_k x_l + \sum_k h_k x_k\right)\right] = \frac{\beta}{Z}\sum_x x_j \exp\left[\beta\left(\sum_{k,l}J_{kl}x_k x_l + \sum_k h_k x_k\right)\right] \quad \text{(S6)}$$

$$= \beta \sum_x x_j P(x) = \beta \langle x_j \rangle,$$

and so

$$\frac{1}{Z^2}\frac{\partial Z}{\partial h_j}\sum_x x_i \exp\left[\beta\left(\sum_{k,l}J_{kl}x_k x_l + \sum_k h_k x_k\right)\right] = \frac{\beta\langle x_j\rangle}{Z}\sum_x x_i \exp\left[\beta\left(\sum_{k,l}J_{kl}x_k x_l + \sum_k h_k x_k\right)\right] \quad \text{(S7)}$$

$$= \beta \langle x_i \rangle \langle x_j \rangle.$$

Thus, putting terms together, we have

$$\chi_{ij} = \frac{\partial \langle x_i \rangle}{\partial h_j} = \beta\big(\langle x_i x_j \rangle - \langle x_i \rangle \langle x_j \rangle\big). \quad \text{(S8)}$$

We therefore find that the average response of component $i$ to a perturbation on component $j$ is equal to the spontaneous equilibrium correlation between $i$ and $j$ (scaled by the inverse temperature $\beta$). This is precisely the fluctuation-dissipation theorem for the equilibrium Ising model.